\documentclass[a4paper]{article}

\usepackage{INTERSPEECH2020}
\usepackage[acronym, nomain, nopostdot, nonumberlist, toc]{glossaries}
\usepackage[colorlinks=true]{hyperref}
\hypersetup{ %
    colorlinks=true,
    linkcolor=black,
    citecolor=black,
    filecolor=black,
    urlcolor=black
}
\usepackage{url}
\usepackage{balance}

\usepackage[capitalize,noabbrev]{cleveref}
\usepackage{subcaption}

\usepackage{etoolbox}
\newcommand{\zerodisplayskips}{%
  \setlength{\abovedisplayskip}{0pt}%
  \setlength{\belowdisplayskip}{0pt}%
  \setlength{\abovedisplayshortskip}{0pt}%
  \setlength{\belowdisplayshortskip}{0pt}%
  \setlength{\abovecaptionskip}{0pt}%
  \setlength{\belowcaptionskip}{0pt}
}
\appto{\normalsize}{\zerodisplayskips}
\appto{\small}{\zerodisplayskips}
\appto{\footnotesize}{\zerodisplayskips}

\newacronym{ae}{AE}{AutoEncoder}
\newacronym{gan}{GAN}{Generative Adversarial Network}
\newacronym{vc}{VC}{Voice Conversion}
\newacronym{mse}{MSE}{Mean Squared Error}
\newacronym{cnn}{CNN}{Convolutional Neural Network}
\newacronym{stft}{STFT}{Short-Time Fourier Transform}
\newacronym{cola}{COLA}{Constant OverLap-Add}
\newacronym{asr}{ASR}{Automatic Speech Recognition}
\newacronym{sgd}{SGD}{Stochastic Gradient Descent}
\newacronym{vq}{VQ}{Vector Quantization}
\newacronym{vae}{VAE}{Variational AutoEncoder}
\newacronym{tts}{TTS}{Text-to-Speech}
\newacronym{ppg}{PPG}{Phonetic PosterioGram}
\newacronym{ar}{AR}{Auto-Regressive}
\newacronym{mos}{MOS}{Mean Opinion Score}
\newacronym{ge2e}{GE2E}{Generalized End-to-End}
\newacronym{mfcc}{MFCC}{Mel-Frequency Cepstral Coefficient}
\newacronym{dtw}{DTW}{Dynamic Time Warping}
\newacronym{pca}{PCA}{Principal Component Analysis}
\newacronym{tsne}{t-SNE}{t-Distributed Stochastic Neighbor Embedding}
\newacronym{pllr}{PLLR}{Phonological Log-Likelihood Ratio}
\newacronym{rnn}{RNN}{Recurrent Neural Network}
\newacronym{pesq}{PESQ}{Perceptual Evaluation of Speech Quality}
\newacronym{vad}{VAD}{Voice Activity Detector}
\newacronym{vctk}{VCTK}{Voice Cloning Toolkit}

\usepackage{amssymb,amsmath,amsthm,dsfont}

\usepackage{bbm}

\providecommand{\norm}[1]{\ensuremath{\left\lVert#1\right\rVert}}

  \DeclareMathOperator{\E}{{\mathbb E}}

  \renewcommand{\ss}{\mathbf{s}}

  \providecommand{\mX}{\mathbf{X}}

  \providecommand{\cL}{\mathcal{L}}

\RequirePackage[colorinlistoftodos,bordercolor=orange,backgroundcolor=orange!20,linecolor=orange,textsize=scriptsize]{todonotes}
\providecommand{\comment}[2]{\todo[inline,caption={}]{\textbf{#1: }#2}}%
\providecommand{\inlinecomment}[3]{%
  {\color{#1}#2: #3}}%
\newcommand\commenter[2]%
{%
  \expandafter\newcommand\csname i#1\endcsname[1]{\inlinecomment{#2}{#1}{##1}}
  \expandafter\newcommand\csname #1\endcsname[1]{\comment{#1}{##1}}
}

\title{FastVC: Fast Voice Conversion with non-parallel data}
\name{Oriol Barbany$^{1,2}$, Milos Cernak$^1$}
\address{
  $^1$Logitech Europe S.A., 1015, Lausanne, Switzerland \\
  $^2$\'{E}cole Polytechnique F\'{e}d\'{e}rale de Lausanne (EPFL), 1015, Lausanne, Switzerland}
\email{milos.cernak@ieee.org}

\begin{document}

    \maketitle
    \begin{abstract}
        This paper introduces FastVC, an end-to-end model for fast \gls{vc}. The proposed model can convert speech of arbitrary length from multiple source speakers to multiple target speakers. FastVC is based on a conditional \gls{ae} trained on non-parallel data and requires no annotations at all. This model's latent representation is shown to be speaker-independent and similar to phonemes, which is a desirable feature for \gls{vc} systems. While the current \gls{vc} systems primarily focus on achieving the highest overall speech quality, this paper tries to balance the development concerning resources needed to run the systems. Despite the simple structure of the proposed model, it outperforms the \gls{vc} Challenge 2020 baselines on the cross-lingual task in terms of naturalness.
    \end{abstract}
    
    
    \section{Introduction}
    
    The \gls{vc} task consists of modifying a speech signal uttered by some source speaker as another target speaker uttered it. Cross-lingual \gls{vc} allows using different source and target languages. In such a task, the linguistic information is preserved, but speaker-dependent features are changed, which requires semantic reasoning about the input signal.
    
    \Gls{vc} is an inherently ill-posed problem; there are multiple correct outputs. Even if it is easy for humans to identify the concepts of content and style (assessed as naturalness and speaker similarity), it is difficult to quantify the conversion's overall speech quality. On the one hand, the lack of objective measures hinders choosing a training strategy and an objective function. On the other hand, most of the works in \gls{vc} only report subjective scores. Some works on parallel \gls{vc} report objective metrics such as the Root Mean Square Error (RMSE) \cite{rmse}, but those are not always correlated with human perception \cite{sisman2020overview}. This is especially the case for non-parallel scenarios, where even \gls{pesq}, which predicts the human-perceived speech quality with respect to a target signal, does not correlate with subjective scores (see \cref{sec:objective}).
    
    The subjective score indicates how a system works and may be biased depending on the test setup. Moreover, it is impossible to compare available systems evaluated only by different evaluators' subjective tests and under different conditions. The \gls{vc} Challenge\footnote{\url{http://www.vc-challenge.org/}} circumvents this problem by providing a common evaluation dataset and performing large-scale crowd-sourced perceptual evaluations.

    \gls{vc} requires the factorization of speech into linguistic and non-linguistic information. Therefore, one natural approach is to use representations that lack speaker information, which are well studied in the speech recognition domain. Such representations are then concatenated with the information of the target speaker (studied in the speaker identification and verification domains) and mapped to the waveform domain yielding the \gls{vc} output. One example of such an approach achieving high-quality conversions consists of first transcribing audio to text using an \gls{asr} system and then using a \gls{tts} model conditioned on the target speaker and the obtained transcription \cite{hayashi2019espnettts, inaguma2020espnet, watanabe2018espnet}.

    Another common approach when dealing with non-parallel data is to train an \gls{ae} model in speech reconstruction and enforce speaker independence in the latent representations. Such latent features are then used in the same fashion as the speaker-independent representations of the former approach. This method can suppress the need for annotated data and find representations that potentially preserve para-linguistical information.
    
    In line with the \gls{vc} Challenge 2020, this work uses non-parallel data for the \gls{vc} model training. In particular, the proposed method is based on conditional \glspl{ae} as in AutoVC~\cite{autovc}, but with focus on fast conversions.
    
    \section{Related Works}
    
    The approaches using speaker-independent representations leverage pre-trained models for its computation. In general, these systems require large amounts of transcribed data, whose collection is very costly and time-consuming \cite{Pascual2019}. Moreover, such speaker-independent features usually lack para-linguistical information such as the intonation, which can potentially lead to conversions having different meanings in some cases. In the case of using text, the timing information is also lost, and an additional mapping to phonetic transcriptions is needed in cross-lingual settings.
    
    Speaker independence can be explicitly enforced with an adversarial setting, where a classifier is trained to predict the speaker from the latent representation. The loss from such classifiers can then be used to learn the mapping from the input signal to the latent space \cite{chou2018multitarget,starganvc}.
    
    One can also implicitly enforce speaker independence by reconstructing the speech from the low-dimensional latent representation and uncompressed speaker information. Speaker disentanglement, in this case, follows from the redundancy principle \cite{barlow-redundancy}. Since the waveform generator is explicitly conditioned on the speaker identity, the feature extractor does not have to capture speaker-dependent information in the latent features. The desired speaker independence is achieved if the dimension of the latent space satisfies the following trade-off. On the one hand, it has to be sufficiently small to factor out the speaker's information. On the other hand, it has to be large enough to allow for perfect reconstruction and capture as much of the input data as possible.
    
    AutoVC achieves a speaker-independent latent space by implementing the previous approach with a simple conditional \gls{ae}. The encoder network of the \gls{ae} applies an information bottleneck that both reduces the number of features and downsamples the signal in the temporal dimension.
    
    CycleVAE \cite{cyclevae} models the information bottleneck with a \gls{vae}, which means that the latent features are enforced to follow a known distribution. \gls{vq}-\gls{vae} \cite{vq-vae} uses \gls{vq} as an additional information bottleneck on the latent features obtained with a \gls{vae}. Time-jitter regularization, consisting of replacing each latent vector with either one or both of its neighbors, was also shown to be useful as an additional information bottleneck in top of the former approach~\cite{unsupervised-representation}. This regularization helps to model the slowly-changing phonetic content by avoiding the use of latent vectors as individual units. The authors claimed that the latent representations found by \glspl{ae} do not factor the speaker out. Instead, their findings showed that a \gls{vae} or its \gls{vq} version was required to achieve speaker disentanglement. However, \cite{spk-encoder} showed that conditional \glspl{ae} with speaker-dependent encoders effectively achieve the desired speaker-independent latent representations.
    
    \section{FastVC}
    
\begin{figure*}[ht]
        \centering
        \includegraphics[width=.9\textwidth]{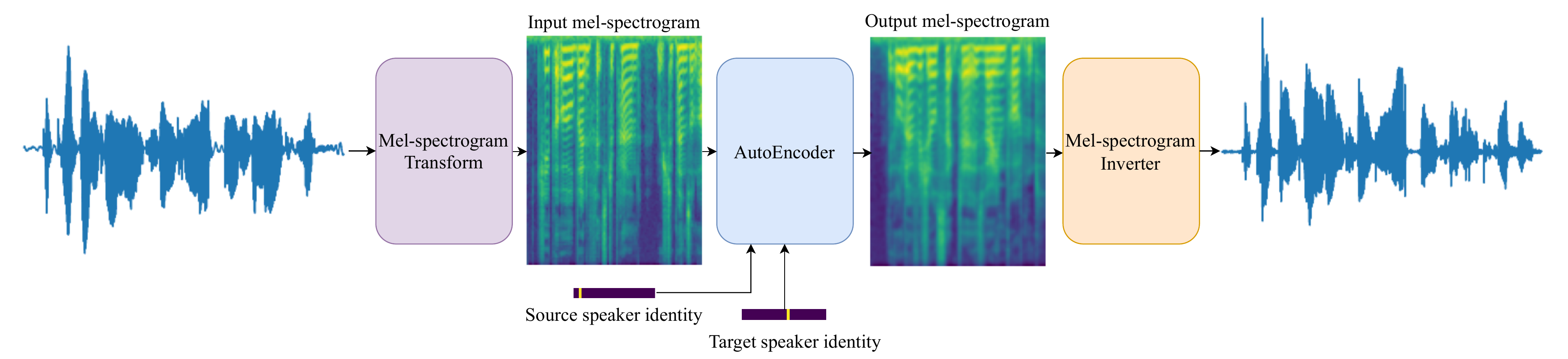}
        \caption[FastVC model architecture.]{FastVC model architecture during conversion mode. During training, both the the source and target speaker identities are the same.}
        \label{fig:fastvc}
    \end{figure*}

\begin{figure*}[ht]
        \centering
        \includegraphics[width=.8\textwidth]{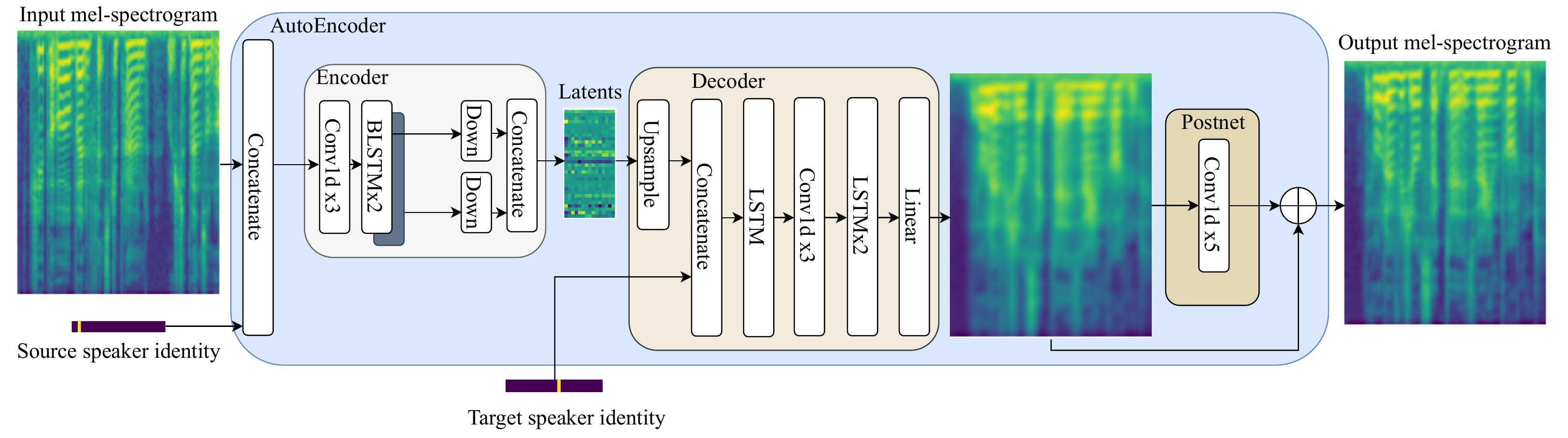}
        \caption[AutoEncoder module of FastVC.]{Diagram of the \gls{ae} module for FastVC during conversion mode. The \texttt{AutoEncoder} comprises the \texttt{Encoder} and the \texttt{Decoder}, but also the \texttt{PostNet}. The \texttt{PostNet}, which is proposed in \cite{autovc}, builds the finer details of the spectrogram, which is excessively smooth before this module.}
        \label{fig:fastvc_autoencoder}
    \end{figure*}

    FastVC is an end-to-end model that performs fast many-to-many \gls{vc} and is trained using non-parallel data. This system performs \gls{vc} by learning a mapping between the source and converted waveforms. This latter has the same linguistic information as the source speech but different speaker information. In particular, FastVC learns this mapping with a conditional \gls{ae} framework that is trained on the reconstruction of Mel-spectrograms similarly to \cite{autovc}. FastVC performs \gls{vc} with the three-stage model depicted in \cref{fig:fastvc}. Its \gls{ae} module is depicted in \cref{fig:fastvc_autoencoder}.

    Equally to AutoVC~\cite{autovc}, FastVC uses log-scale Mel-spectrograms with 80 Mel channels as inputs. However, the Mel-spectrogram module is a~\gls{cnn}-based learnable module and not a fixed transformation as in AutoVC. This module can be trained if desired and is initialized to provide exact Mel-spectrograms. This allows using raw speech waveforms as input instead of Mel-spectrograms.
    
    The theoretical guarantees justifying the \gls{vc} capabilities of \cite{autovc} hold under the assumption that the speaker embeddings of different utterances of the same speaker are the same, and those from different speakers are distinct. In order to satisfy this assumption, FastVC uses one-hot encoded speaker embeddings similarly to \cite{vq-vae}.
    
    In FastVC, both the encoder $E(\cdot,\cdot)$ and the decoder $D(\cdot,\cdot)$ are conditioned on the speaker identity of the source and target speaker as in~\cite{spk-encoder}. This speaker identity is concatenated with the other input signal at every time step. One of the most key design choices in implementing a conversion function for \gls{vc} learned on speech reconstruction is choosing adequate information bottlenecks.
    
    FastVC uses dimensionality reduction in the frequency dimension and temporal downsampling as in \cite{autovc}. The latent features are then upsampled to match the original time rate using a causal variant of the nearest neighbor interpolation technique. This can be seen as a causal version of the time-jitter regularization proposed in \cite{unsupervised-representation}, with the time jitter as a hyper-parameter that corresponds to the downsampling factor.
    
    The information bottleneck introduces two pivotal hyperparameters for the speaker disentanglement; the latent features' dimension and the downsampling factor. In particular, FastVC doubles the temporal downsampling factor with respect to \cite{autovc}. This design choice achieves speaker-independent phoneme-like latent features that solve the pitch inconsistency problems of AutoVC reported in \cite{autovc2}. For more details on this, refer to \cite{barbany_msc}.
     
    FastVC generates speech with a sampling rate of 22050 Hz, which makes \gls{ar} models unsuitable, especially if fast conversions are desired. \cite{autovc} used WaveNet~\cite{wavenet} conditioned on the log-scaled Mel-spectrogram as a generative model for raw speech. To achieve fast inference, FastVC resorts to using a non-\gls{ar} generative model. This design choice is the main reason for the fast conversions obtained with this approach. In particular, the Mel-spectrogram inverter is chosen to be MelGAN, introduced by \cite{melgan}.
    
    \section{Experimental setup}
    
    \subsection{Datasets}
    
    FastVC only trains on raw speech waveforms and speaker identities. This means that it does not requires any additional annotation. The main dataset used for this project is the \gls{vctk} described in \cite{vctk}. The \gls{vctk} dataset is chosen as the main dataset for its widespread use for the \gls{vc} task \cite{vq-vae,chou2018multitarget,autovc,autovc2}.
    
    The model comparison of the \gls{vc} Challenge is performed with samples generated using the dataset of the same Challenge. The use of the Challenge training dataset is essential but not enough to train a model such as FastVC. In this project, the \gls{vctk} and \gls{vc} Challenge datasets were simply merged, which is allowed in the Challenge.
    
    The TIMIT dataset \cite{timit} is chosen for the latent features' analysis (see \cref{sec:latents}). This dataset is public and contains speech data and hand-verified time-aligned phoneme transcriptions. In particular, only the test partition of this dataset is used to avoid incorporating many new speakers.
    
    \subsection{Training}
    
    \glspl{gan} are notoriously difficult to train, with mode collapse and oscillations being a common problem \cite{liang2018generative}. For this reason, the basic FastVC model (denoted FastVC in \cref{tab:pesq}) uses the pre-trained weights for MelGAN provided by \cite{melgan}. The \gls{ae} module in FastVC is trained from scratch to match the conditioning signal required by the Mel-spectrogram inverter. The basic version of FastVC is obtained by only training the \gls{ae} module using the ADAM optimizer \cite{adam} with a learning rate of 0.001, $\beta_1=0.9$, and $\beta_2=0.99$ for 200 epochs. The training objective for this setting is presented in \eqref{eq:basic_loss}, where $\mX$ is the input Mel-spectrogram, $\ss$ its correspondent speaker and $\hat{\mX}$ the \gls{ae} output.
    
    \begin{align}
        \cL_{content} = \norm{E(\mX,\ss) - E(\widehat{\mX},\ss)}^2
        \label{eq:content}
    \end{align}
    
    \begin{align}
        \min \E_{\mX,\ss}\big[&\norm{\mX-\widehat{\mX}}^2 + \norm{\mX-D(E(\mX,\ss),\ss)}^2 +\cL_{content} \big]
        \label{eq:basic_loss}
    \end{align}
    
    \emph{FastVC with end-to-end training}: to allow the model to use information that may be not included in the Mel-spectrogram and generate more efficient representations for the task of \gls{vc}, FastVC also allows end-to-end training. The weights obtained in the only-\gls{ae} training are used as a starting point. In this setup, FastVC behaves as the generator of a \gls{gan}, which takes raw speech as input.
    
    FastVC uses multiple discriminators that run at different rates, as proposed in \cite{melgan}. To ensure that the linguistic information is captured, the content loss term \eqref{eq:content} is added to the generator objective in \cite{melgan}. The content loss term enforces the codes of the original and converted speech to be the same, i.e., it enforces \gls{vc} to be idempotent. We speculate that this is enough to achieve quality speech that preserves the lexical content.
    
    The content loss is weighted by a factor of $20$ and added to the generator's total objective. The regularization amount is chosen to ensure that the losses have the same order of magnitude, and they both decrease individually. For this setting, ADAM is also used as the optimization algorithm. In this case, however, with a learning rate of $10^{-4}$, $\beta_1=0.5$, and $\beta_2=0.9$ for 200 epochs. These specific values are suggested in \cite{gan_training} to train \glspl{gan} with ADAM, and also used in \cite{melgan}.
    
    All the experiments use a batch size of 16, and the merged dataset is randomly split into 90\% for training and 10\% for testing purposes. During training, FastVC is fed chunks of 8192 samples of speech sampled at 22050 Hz, while in inference, the model's input is the whole waveform.
    
    \section{Results}
    
    FastVC converts voices $4\times$ faster than real-time, and $500\times$ faster than AutoVC, measured on Intel(R) Core(TM) i7-8700K at 3.70GHz.
    
    \subsection{Objective assessment}
    \label{sec:objective}
    
    One of the main difficulties in building \gls{vc} models is that there are no standardized objective measures. The lack of such metrics hinders the system comparison and the performance of ablation studies. The variant of FastVC submitted to the \gls{vc} Challenge was chosen based on the value given by \gls{pesq}, an objective method that rates the speech quality by predicting the \gls{mos}.
    
    The fact that \gls{pesq} is not used as a standardized measure to substitute \gls{mos} is that the former requires both the desired waveform and the one generated with the evaluated system. The approach that FastVC takes to deal with non-parallel data is to learn the conversion function on the task of speech reconstruction. In this case, the \gls{pesq} measure is more suited since self-reconstruction was learned during training, and mapping to the same speaker is a valid \gls{vc} instance.
    
    \cref{tab:pesq} shows the obtained results. Note that the reconstruction performance alone is not a useful metric to evaluate a \gls{vc} system because it does not measure the speaker's disentanglement. In particular, perfect reconstruction can be achieved if there is no information bottleneck and thus the latent features are not speaker-independent. Therefore, this metric should be used for systems with speaker-independent inputs.
    
    \begin{table}[t]
        \centering
        \begin{tabular}{p{5.4cm}p{1.6cm}}
            \toprule
            Experiment & PESQ\\
            \midrule
            AutoVC -- baseline \cite{autovc} & $\mathbf{2.56 \pm 0.23}$ \\            
            \midrule
            FastVC with information bottleneck proposed in~\cite{autovc} & $2.57 \pm 0.25$\\
            FastVC with 10 Hz latent features & $2.61 \pm 0.24$\\
            FastVC with adversarial speaker classifier & $2.62 \pm 0.24$\\
            FastVC (VCC20 submission) & $\mathbf{2.68 \pm 0.22}$\\
            \midrule
            FastVC with end-to-end training & $1.56 \pm 0.29$ \\
            FastVC with learnable Mel-spectrogram & $1.50 \pm 0.40$ \\
            PhonetVC (\cref{sec:phonetvc}) & $\mathbf{1.67 \pm 0.30}$\\
            \bottomrule
        \end{tabular}
        \caption{Objective results performed over 100 utterances of less than 5 seconds from the test partition. The reported values are the mean and the standard deviation of the sample. First three FastVC variants are described later in \cref{sec:latents}.}
        \label{tab:pesq}
        \vspace{-2em}
    \end{table}
    
    FastVC with end-to-end training performs worse in terms of \gls{pesq} than FastVC. This can be justified because, in the end-to-end training, the aim is not to match the input Mel-spectrogram but to maximize the \gls{gan} objective. A future subjective evaluation would be needed to confirm if the \gls{pesq} also correlates with the perceived quality in such cases.
    
    The use of this metric with parallel utterances aligned using \gls{dtw} was also explored. However, in this case, the results were inconclusive and not related at all with perceptual scores. The ill-posedness of the problem can justify this; a sound output other than the time-aligned parallel utterance (or the input utterance in the evaluation of reconstruction, especially on the end-to-end case) may be obtained.
    
    \subsection{Subjective assessment}
    The subjective scores for the cross-lingual \gls{vc} task are presented in the \gls{vc} Challenge 2020 paper. FastVC is represented with the label \textbf{T15}. You can also compare the \gls{vc} Challenge baselines, AutoVC, and the proposed FastVC at \url{https://barbany.github.io/fast-vc/}.
    
    \subsection{Latent space analysis}
    \label{sec:latents}
    
    \subsubsection{Speaker independence}
    
    Prosodic information leaks through the bottleneck of AutoVC, causing the target pitch to fluctuate unnaturally~\cite{autovc2}. To tackle this issue, the authors proposed in \cite{autovc} to remove the speaker identity from the latent representations and the prosodic information. The temporal downsampling factor proposed in \cite{autovc2} matches the design choice of FastVC. With this value, FastVC outputs do not have the unnatural pitch jumps of AutoVC without the need of disentangling the prosodic information from the latent features and introducing the synthetic target prosody. Refer to \cite{barbany_msc} for more details.
    
    FastVC requires that the latent features are speaker-independent, but this is not explicitly enforced. The fact that the encoder disentangles the speaker in an unsupervised fashion can be explained with the redundancy principle \cite{barlow-redundancy}. However, adversarial training of the latent representations as in \cite{chou2018multitarget} could further disentangle the speaker's information and downplay the information bottleneck's design choices.
    
    To confirm if the redundancy principle suffices, a variant of FastVC with an adversarial speaker classifier was implemented. In particular, an adaptation of the discriminator used to achieve class-independent latent representations in \cite{musictranslation} is implemented. The minimax game here is for the encoder to seek class-independent latent features and the classifier to classify them correctly. The classifier is trained with the cross-entropy loss on the speaker labels using the ADAM optimizer with a learning rate of 0.001, $\beta_1=0.9$, and $\beta_2=0.99$. The negative loss, termed as domain confusion loss in \cite{musictranslation}, is added as a regularizer to \eqref{eq:basic_loss} with a weighting of $0.1$ so that each individual objective had the same order of magnitude.
    
    Even if the classifier network was trained simultaneously as FastVC, the prediction accuracy was 0\% when the speaker-independence signal was used with the latent features and when it was not. These results were obtained with a model trained using the 278 speakers resulting from the mix of the \gls{vctk} corpus and the test partition of the TIMIT dataset. The speaker-independence results suggest that the redundancy principle is enough to achieve speaker-independence, which is in line with the results reported in \cite{autovc,autovc2}.
    
    \subsubsection{Phonetic similarity}
    
    Similarly to \cite{vq-vae,unsupervised-representation}, the latent features of FastVC lack speaker information and are potentially similar to phonemes. A perceptron is used to find a hypothetically simple correspondence between phonemes and the latent features. This model is trained using the latent representations extracted from the TIMIT test data with a trained FastVC network. The obtained latent features are randomly split into the train (70\%), validation (10\%), and test (20\%) sets.
    
    The information bottleneck on the temporal dimension of FastVC yields a latent representation with a 2.5 Hz rate. This rate is a factor of 10 lower than the rate of the latent features in \cite{vq-vae}. The average phoneme rate is around 10 Hz \cite{inforate_speech,phon_rate}, which means that each latent vector at a given time represents more than one phoneme. For the classification task, each latent vector was assumed to represent the phoneme with a larger intersection in the temporal domain.
    
    The phoneme classifier was trained by minimizing the cross-entropy loss with \gls{sgd} and early stopping on the loss on the validation partition of the dataset containing the latent features from the TIMIT test data. This classifier correctly classified 42.45\% of the latent features. In contrast, a random classifier and a classifier always choosing the prior most likely phoneme on the train partition had an accuracy of 2.44\% and 9.43\%. These results suggest that there is indeed a correspondence between the latent features and phonemes. For comparison, \gls{vq}-\gls{vae} \cite{vq-vae} uses a 128-dimensional discrete space and obtains a classification accuracy of 49.3\%, while choosing the prior most likely phoneme gives a 7.2\%. A classification drop from the results in \cite{vq-vae} is expected due to the lower rate representation and the classifier's simplicity.
    
    Even if the latent representations' low rate suggested that a latent vector represents a combination of sounds rather than a single phoneme, the number of distinct units with groups of phonemes exponentially grows with the group size. This growth implies that there are more classes to predict, and some may not even be seen during training.
    
    \subsection{PhonetVC}
    \label{sec:phonetvc}
    
    PhonetVC is a variant of the proposed model designed to confirm the benefits of using the latent features obtained by FastVC instead of speaker-independent speech features. PhonetVC uses an estimation of the \gls{pllr} features computed with Phonet~\cite{phonet} instead of the latent representations obtained by the encoder network in \cref{fig:fastvc_autoencoder}. The resulting model works with speech at 16 kHz, and the Decoder, Postnet, and Mel-spectrogram inverter are jointly trained from scratch using the MelGAN training objective \cite{melgan}.
    
    \section{Conclusions}
    
    This work proposed a fast and competitive \gls{vc} system. It is worse at capturing the speaker's style of speakers with little data in comparison to its quality performance (see subjective results in the \gls{vc} Challenge paper). This is justified by the fact that the training dataset is very imbalanced concerning the language, and the performance could be degraded for non-English speakers. One possible approach to tackle the language imbalance problem is to incorporate additional non-English speech datasets to balance the languages. However, the percentage of data per speaker on the \gls{vc} Challenge would be even smaller in this case. A different approach to tackle dataset imbalance is the multi-reader technique described in \cite{dvector}.
    
    \section{Acknowledgements}
    
    We thank Kaizhi Qian for providing the non open-sourced full code of AutoVC used as a starting point for this project.
    \balance
    \bibliographystyle{IEEEtran}
    \bibliography{thesis.bib}
\end{document}